\documentclass[revtex,aps,prb,reprint,superscriptaddress,amsmath,amssymb,showkeys,showpacs,preprintnumbers]{revtex4-1}

\usepackage{graphicx}
\usepackage{dcolumn}
\usepackage{bm}
\usepackage{float}

\begin{document}

\title{Spin-torque efficiency enhanced by Rashba spin splitting in three dimensions}

\author{Kazuhiro Tsutsui}\email[]{tsutsui@stat.phys.titech.ac.jp}
\affiliation{Department of Physics, Tokyo Institute of Technology, 2-12-1 Ookayama, Meguro-ku, Tokyo 152-8551, Japan}

\author{Shuichi Murakami}\email[]{murakami@stat.phys.titech.ac.jp}
\affiliation{Department of Physics, Tokyo Institute of Technology, 2-12-1 Ookayama, Meguro-ku, Tokyo 152-8551, Japan}

\date{\today}

\begin{abstract}
We examine a spin torque induced by the Rashba spin-orbit coupling in three dimensions within the Boltzmann transport theory.
We analytically calculate the spin torque and show how its behavior is related with the spin topology in the Fermi surfaces by studying the Fermi-energy dependence of the spin torque.
Moreover we discuss the spin-torque efficiency which is the spin torque divided by the applied electric current in association with the current-induced magnetization reversal.
It is found that high spin-torque efficiency is achieved when the Fermi energy lies on only the lower band and there exists an optimal value for the Rashba parameter, where the spin-torque efficiency becomes maximum.
\end{abstract}


\maketitle

\section{Introduction}

One of the purposes of spintronics is to control the magnetization direction of a ferromagnet by using electric currents instead of an external magnetic field.
Recently the magnetization reversal due to spin-transfer-torque effect \cite{Slonczewski96, Berger96} has been intensively investigated, which requires multilayer structures such as spin valves, tunnel junctions or domain walls.
On the other hand, another method to switch magnetization direction, which is due to the spin-orbit coupling (SOC), was suggested theoretically \cite{Tan07, Obata08, Manchon08, Manchon09, Matos-Abiague09, Tan11} and verified experimentally \cite{Chernyshov09, Miron10}.
For instance, a giant spin torque is observed in an asymmetric ferromagnetic metal layer AlO${}_x$/Co/Pt, and a current-driven magnetic field of $1$ T for a driven current $10^8$ A/cm${}^2$ is reported \cite{Miron10}.

In the above systems, the Rashba-type spin splitting is dominant owing to the interplay between the asymmetric structure and a strong SOC derived from the Pt atom.
Rashba systems \cite{Rashba60} under an external electric field or a current injection become spin-polarized because the spin distribution on the Fermi surfaces becomes imbalanced.
In ferromagnetic metals, on the other hand, there exists an exchange coupling between conduction electrons spins and localized spins.
Under the non-equilibrium state, magnetization in ferromagnetic Rashba systems is macroscopically given a spin torque.
Now if we assume a single-domain ferromagnet, the stability of magnetic ordering is characterized by the anisotropic field.
Thus the magnetization can be reversed when the spin torque overcomes the anisotropic field.
Contrary to a spin-transfer torque, there is no transfer of spin angular momentum from outside.
Rather, orbital angular momentum in a crystal is converted to spin angular momentum via the spin-orbit coupling, and is transfered to a ferromagnet.
This mechanism is intrinsic to the band structure and does not require two non-collinear ferromagnets.

Moreover, it is theoretically reported that compared with some systems with the spin-orbit coupling due to impurities or Luttinger spin-orbit bands, the Rashba system presents a giant spin torque to reverse the magnetization owing to the inversion asymmetry \cite{Manchon09}.
Therefore, a spin torque due to the Rashba spin-orbit coupling attracts many interests as a realistic candidate in spintronic applications.

In this paper, we explore possibility of the current-induced magnetization reversal by examining three-dimensional models with the Rashba SOC (3D Rashba models) theoretically, and compare the results with the 2D Rashba models.
We focus on the low-density regime where only the lower band lies on the Fermi energy.
This low-density regime has not been studied for 2D Rashba models, either.
This low-density regime becomes realistic when the Rashba parameter is large, e.g. in the recent discovery of the new bulk Rashba semiconductor BiTeI \cite{Ishizaka11,Bahramy11}.
In this material, the Rashba effect is induced by the structural inversion asymmetry in the bulk crystal structure, and therefore the Rashba SOC is much stronger than the typical value of the Rashba SOC in 2D semiconductor heterostructures \cite{Nitta97} or that in metal surfaces \cite{LaShell96, Ast07}.

As we vary the Fermi energy, the topology of the Fermi surface changes.
Correspondingly, we found that the spin torque as a function of the Fermi energy $E_F$ behaves differently between the both sides of the topological transition.
Moreover, we examine the spin-torque efficiency which is the spin torque divided by the applied electric current, in order to discuss how to enhance the efficiency in association with the current-induced magnetization reversal.

\section{3D Rashba model}

\begin{figure}
 \begin{center} 
 \includegraphics[width=80mm]{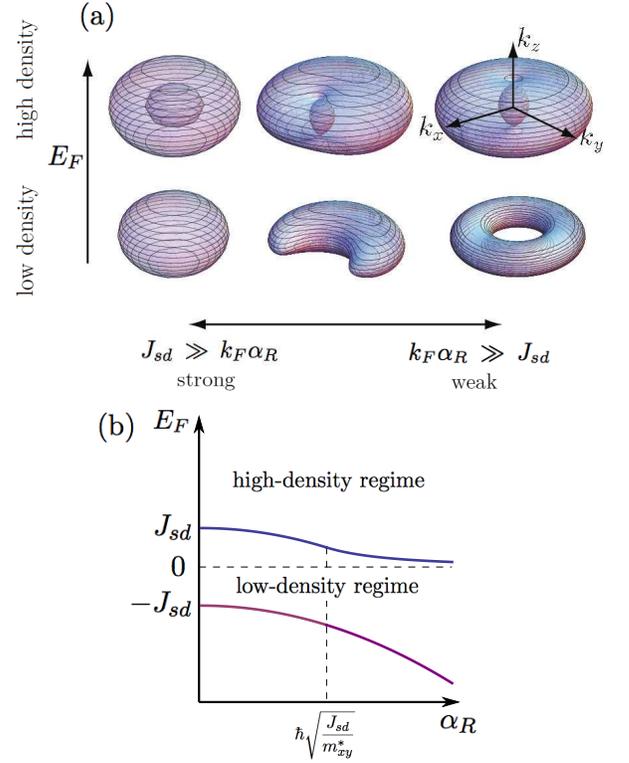} 
 \end{center} 
 \caption{(Color online) Fermi surfaces in the 3D Rashba model and their topological change. (a) Schematic figures of Fermi surfaces in the limit of strong ($J_{sd}\gg k_F\alpha_R$), intermediate and weak exchange coupling regime ($k_F\alpha_R\gg J_{sd}$), and in high and low-density regimes.
 (b) Phase diagram of the high- and low-density regimes as a function of $\alpha_R$.
 } 
 \label{FS} 
 \end{figure}

We calculate the spin torque on the magnetization of a ferromagnet driven by the spin polarization of conduction electrons in systems with a strong SOC.
We consider three-dimensional models with the Rashba effect, i.e., 3D Rashba models \cite{Ishizaka11, Bahramy11}.
We take the direction of structural inversion symmetry breaking as $z$ axis and conduction electrons move in three dimensions.
In our model, we also include localized spins coupled to the conduction electrons via the exchange coupling.
Our Hamiltonian is thus described by
\begin{equation}
{\cal H}=\frac{\hbar^2}{2m_{xy}^*}(k_x^2+k_y^2)+\frac{\hbar^2}{2m_{z}^*}k_z^2+\alpha_R\bm{e}_z\cdot (\bm{k}\times \bm{\sigma})-J_{sd}\bm{M}\cdot \bm{\sigma},
\end{equation}
where $m_{xy}^*, m_z^*$ represent the effective masses of conduction electrons with the $xy$ plane and along the $z$ axis respectively, $\alpha_R$ represents the Rashba parameter, $J_{sd}$ is an exchange coupling between conduction electrons and magnetization, $\bm{M}=(\cos\varphi_M,\sin\varphi_M)$ is the direction of magnetization and $\bm{\sigma}=(\sigma_x, \sigma_y, \sigma_z)$ is the Pauli matrices.
$\bm{e}_z$ denotes the unit vector in the $z$ direction.
The eigenenergies and eigenstates of the above Hamiltonian are respectively given by
\begin{eqnarray}
E_s(\bm{k}) &=& \frac{\hbar^2}{2m_{xy}^*}(k_x^2+k_y^2)+\frac{\hbar^2}{2m_{z}^*}k_z^2 \nonumber \\
 &+& s \sqrt{(\alpha_R k_y-J_{sd}\cos\varphi_M)^2+(\alpha_R k_x+J_{sd}\sin\varphi_M)^2}, \label{Es} \nonumber \\
 & & \\
\Psi_{\bm{k},s} &=& \frac{1}{\sqrt{2}}\left(
\begin{array}{c}
se^{i\gamma_{\bm{k}}}  \\
1
\end{array}
\right)e^{i\bm{k}\cdot \bm{r}}, \label{wf}
\end{eqnarray}
where $s$ is a band index with $+1$ for the upper band and $-1$ for the lower band and $\tan\gamma_{\bm{k}}\equiv \frac{\alpha_R k_x+J_{sd}\sin\varphi_M}{\alpha_R k_y-J_{sd}\cos\varphi_M}$.
Equations (\ref{Es}) and (\ref{wf}) look similar to 2D Rashba models, but the shape of Fermi surfaces is nontrivial in 3D Rashba models as shown in Fig. \ref{FS} (a).

In the regime of strong exchange coupling, i.e., $J_{sd}\gg k_F\alpha_R$, the Fermi surfaces are mainly governed by the Zeeman splitting, where $k_F$ denotes the Fermi wave number.
When the Fermi energy crosses both the upper and lower bands (which we refer to as the high-density regime), the system has a larger and a smaller ellipsoidal Fermi surfaces.
When the Fermi energy lies on only the lower band (which we refer to as the low-density regime), the system has a single ellipsoidal Fermi surface.
In the regime of weak exchange coupling, i.e., $k_F\alpha_R\gg J_{sd}$, on the other hand, the Fermi surfaces are mainly determined by the Rashba-type spin splitting.
In the high-density regime, an apple-like and lemon-like Fermi surfaces are obtained.
In the low-density regime, we obtain a donuts-like Fermi surface and the distribution of the spin density in the wave-number space is along the azimuthal direction in the $xy$ plane.
From Fig. \ref{FS} (a), we find that there exists a topological transition of the Fermi surfaces, namely, a Lifshitz transition \cite{Lifshitz60} in an intermediate value of the exchange coupling.
In the low-density regime, as $J_{sd}/\alpha_R$ grows, the topology of the Fermi surface changes from the torus $T^2$ to the sphere $S^2$.
We remark that the topological transition occurs on the curve $E=J_{sd}-m^*_{xy}\alpha_R^2/(2\hbar^2)\ (\alpha_R<\hbar \sqrt{J_{sd}/m^*_{xy}}),\ =\hbar^2 J_{sd}^2/(2m^*_{xy} \alpha_R^2)\ (\alpha_R\geq \hbar \sqrt{J_{sd}/m^*_{xy}})$, and the band bottom is on the curve $E=-J_{sd}-m^*_{xy}\alpha_R^2/(2\hbar^2)$ as shown in Fig. \ref{FS} (b).

\section{Spin torque}

From the Heisenberg equation of motion for the conduction-electron spin, the spin-continuity equation is deduced:
\begin{equation}
\frac{d \langle \bm{s}\rangle}{dt}+\nabla\cdot {\cal J}_s=-\frac{J_{sd}}{\hbar} \bm{M}\times \langle \bm{s}\rangle+\frac{\alpha_R}{\hbar} \langle (\bm{k}\times \bm{\sigma})\times \bm{e}_z \rangle, \label{sce}
\end{equation}
where $\bm{s}$ refers to the spin-density operator, ${\cal J}_s$ refers to the spin current tensor and $\langle\cdots \rangle$ denotes the quantum average.
The first term of the right hand in Eq. (\ref{sce}) means the current-induced spin torque to the magnetization, and is denoted by $\bm{T}$.
The second term of the right hand, on the other hand, means the torque due to an effective magnetic field introduced by the Rashba SOC.
We calculate the spin polarization and the electric current of conduction electrons under an electric field using the Boltzmann equation of transport
\begin{equation}
-\frac{e}{\hbar}\bm{E}\cdot \frac{\partial f^0_{\bm{k},s}}{\partial \bm{k}}=\sum_{\bm{k}',s'} W^{s s'}_{\bm{k} \bm{k}'} (f_{\bm{k},s}-f_{\bm{k}',s'}), \label{Blz1}
\end{equation}
where $e$ represents the electric charge, $\bm{E}=(E\sin\theta_E\cos\varphi_E,E\sin\theta_E\sin\varphi_E,E\cos\theta_E)$ is an external electric field, and $f^0_{\bm{k},s}=1/(e^{\beta(E_s-\mu)}+1)$ is the Fermi distribution function with band index $s$.
Within the Boltzmann transport theory, the electric current and the spin density respectively read $\bm{j}_{3D}=-\frac{e}{V} \sum_{\bm{k},s=\pm 1} f_{\bm{k},s} \bm{v}_{\bm{k},s}$, where $\hbar \bm{v}_{\bm{k},s}=\frac{\partial E_s}{\partial \bm{k}}=(\frac{\hbar^2}{m^*_{xy}}k_x+s\alpha_R \sin\gamma_{\bm{k}}, \frac{\hbar^2}{m^*_{xy}}k_y+s\alpha_R \cos\gamma_{\bm{k}}, \frac{\hbar^2}{m^*_z}k_z)$ and $\langle \bm{s}\rangle=\frac{1}{V} \sum_{\bm{k},s=\pm 1} f_{\bm{k},s} \bm{s}_{\bm{k},s}$, where $\bm{s}_{\bm{k},s}=\Psi^\dagger_{\bm{k},s} \bm{s} \Psi_{\bm{k},s}=s(\cos\gamma_{\bm{k}},-\sin\gamma_{\bm{k}},0)$.
Under the short-range impurity potential $V(\bm{r})\equiv V\delta(\bm{r})$, the scattering probability reads
\begin{equation}
W^{s s'}_{\bm{k} \bm{k}'}=\frac{\pi n_i}{\hbar} V^2 (1+ss' \cos(\gamma_{\bm{k}}-\gamma_{\bm{k}'})) \delta(E_s(\bm{k})-E_{s'}(\bm{k}')), \label{Blz2}
\end{equation}
where $n_i$ is the impurity concentration.

In this study, we adopt the approximation of a constant relaxation time and discuss effects beyond the present approximation later.
In the following, we consider the two limiting cases of $J_{sd}\gg \alpha_R k_F$ and $\alpha_R k_F \gg J_{sd}$, and calculate the spin torque in high-density and low-density regimes for each limiting case.
We emphasize that in the 2D Rashba models, the Rashba SOC is typically small, and only the high-density regime is considered in general; in contrast, in the 3D Rashba models such as BiTeI, the Rashba SOC is strong and the low-density regime is significant for experiments.

\subsection{Strong exchange coupling regime}

We consider the regime of a strong exchange coupling, i.e., $J_{sd}\gg k_F \alpha_R$.
For simplicity, we retain up to the first order in $k_F\alpha_R /J_{sd}$, that is, $E_s(\bm{k})\simeq \frac{\hbar^2}{2m_{xy}^*}(k_x^2+k_y^2)+\frac{\hbar^2}{2m_{z}^*}k_z^2+s\left( J_{sd}+\frac{1}{2}\alpha_R(k_x\sin\varphi_M-k_y\cos\varphi_M) \right)$.
Up to the first order in $k_F\alpha_R /J_{sd}$, we have
\begin{eqnarray}
\cos\gamma_{\bm{k}} &\simeq& -\cos\varphi_M \nonumber \\
 & & +\frac{\alpha_R}{J_{sd}}(k_y\sin^2\varphi_M+k_x\sin\varphi_M \cos\varphi_M), \\
\sin\gamma_{\bm{k}} &\simeq& \sin\varphi_M \nonumber \\
 & & +\frac{\alpha_R}{J_{sd}}(k_x\cos^2\varphi_M+k_y\sin\varphi_M \cos\varphi_M).
\end{eqnarray}
%
Fermi surfaces change topologically when $E_F=J_{sd}$ as shown in Fig. \ref{zee} (a).

In the high-density regime ($J_{sd}<E_F$), by integrating over the Fermi surfaces, the spin torque and the electric conductivity for the $xy$-plane projective and $z$-axis direction are calculated as
\begin{eqnarray}
\bm{T}_{3D} &=& \frac{\sqrt{2}}{3\pi^2}\frac{e\tau}{\hbar^5}\alpha_R m^*_{xy}\sqrt{m_z^*} \left[ (E_F-J_{sd})^{3/2}-(E_F+J_{sd})^{3/2} \right] \nonumber \\
 & & \times E_{\parallel} \cos(\varphi_M-\varphi_E) \bm{e}_z, \label{hig} \\
\sigma^{\parallel}_{3D} &=& \frac{\sqrt{2}}{3\pi^2} \frac{e^2\tau}{\hbar^3} \sqrt{m_z^*} \left[ (E_F-J_{sd})^{3/2}+(E_F+J_{sd})^{3/2} \right], \\
\sigma^z_{3D} &=& \frac{m_{xy}^*}{m_z^*}\sigma^{\parallel}_{3D},
\end{eqnarray}
where $E_{\parallel}$ represents the external electric field along the $xy$-plane.
In Eq. (\ref{hig}), the first and second terms come from the upper and lower bands respectively.
We can see that these contributions to $\bm{T}_{3D}$ partially cancel each other.
We also calculate the spin torque in 2D Rashba models for a comparison between 3D and 2D Rashba models.
The Hamiltonian in 2D is described by ${\cal H}_{2D}=\frac{\hbar^2}{2m^*}(k_x^2+k_y^2)+\alpha_R\bm{e}_z\cdot (\bm{k}\times \bm{\sigma})-J_{sd}\bm{M}\cdot \bm{\sigma}$.
In a similar way, we obtain the spin torque as $\bm{T}_{2D}=-\frac{1}{\pi}\frac{e\tau E_{\parallel}}{\hbar^4} m^*\alpha_R J_{sd}\cos(\varphi_M-\varphi_E) \bm{e}_z$ and the electric conductivity as $\sigma^{\parallel}_{2D}=\frac{1}{\pi}\frac{e^2\tau}{\hbar^2} E_F$.

In the low-density regime ($-J_{sd}<E_F<J_{sd}$), on the other hand, the electric current and the spin density respectively read $\bm{j}_{3D}=-\frac{e}{V} \sum_{\bm{k}} f_{\bm{k},-1} \bm{v}_{\bm{k},-1}$ and $\langle \bm{s}\rangle=\frac{1}{V} \sum_{\bm{k}} f_{\bm{k},-1} \bm{s}_{\bm{k},-1}$.
With a similar calculation, we obtain
\begin{eqnarray}
\bm{T}_{3D} &=& -\frac{\sqrt{2}}{3\pi^2}\frac{e\tau}{\hbar^5}\alpha_R m^*_{xy}\sqrt{m_z^*} (E_F+J_{sd})^{3/2} \nonumber \\
 & & \times E_{\parallel} \cos(\varphi_M-\varphi_E) \bm{e}_z, \label{Ts1} \\
\sigma^{\parallel}_{3D} &=& \frac{\sqrt{2}}{3\pi^2} \frac{e^2\tau}{\hbar^3} \sqrt{m_z^*} (E_F+J_{sd})^{3/2}, \label{ss1} \\
\sigma^z_{3D} &=& \frac{m_{xy}^*}{m_z^*}\sigma^{\parallel}_{3D}.
\end{eqnarray}
In 2D Rashba models, the spin torque is $\bm{T}_{2D}=-\frac{1}{2\pi}\frac{e\tau E_{\parallel}}{\hbar^4} m^*\alpha_R (E_F+J_{sd})\cos(\varphi_M-\varphi_E) \bm{e}_z$ and the electric conductivity is $\sigma^{\parallel}_{2D}=\frac{1}{2\pi}\frac{e^2\tau}{\hbar^2} (E_F+J_{sd})$.

\subsection{Weak exchange coupling regime}

Let us consider the regime of a weak exchange coupling, i.e., $\alpha_R k_F \gg J_{sd}$.
Up to the zeroth order in $J_{sd}/k_F \alpha_R$, $E_s(\bm{k})\simeq \frac{\hbar^2}{2m^*_{xy}}(k_{\parallel}+sk_0)^2+\frac{\hbar^2}{2m_{z}^*}k_z^2-E_0$, $\cos\gamma_{\bm{k}}\simeq k_y/k_{\parallel}$ and $\sin\gamma_{\bm{k}}\simeq k_x/k_{\parallel}$, where $k_{\parallel}\equiv \sqrt{k_x^2+k_y^2}$, $k_0\equiv \frac{\alpha_R m^*_{xy}}{\hbar^2}$ and $E_0\equiv \frac{\alpha_R^2 m_{xy}^*}{2\hbar^2}$.
The eigenvector given in the present approximation has the same form as that of 3D Rashba Hamiltonian and therefore the spin density of the total Hamiltonian is the spin polarization induced by only 3D Rashba SOC.
As shown in Fig. \ref{rash} (a), the low-density regime is given by $-E_0<E_F<0$ and Fermi surfaces change topologically at $E_F=0$.

In high-density regime ($0<E_F$), the spin torque and the electric conductivity are thus obtained as
\begin{widetext}
\begin{eqnarray}
\bm{T}_{3D} &=& -\frac{1}{2\pi^2}\frac{e\tau}{\hbar^4}\sqrt{m^*_{xy} m_z^*} J_{sd}\left[ \left( E_F+E_0\right) \arcsin \left( \sqrt{\frac{E_0}{E_F+E_0}} \right)+\sqrt{E_0 E_F} \right] E_{\parallel} \cos(\varphi_M-\varphi_E) \bm{e}_z, \\
\sigma^{\parallel}_{3D} &=& \frac{1}{2\pi^2}\frac{e^2\tau}{\hbar^4} \sqrt{m^*_{xy} m_z^*} \alpha_R \left[ (E_F+E_0)\arcsin \left( \sqrt{\frac{E_0}{E_F+E_0}} \right)+\left( \frac{4}{3}E_F+E_0 \right)\sqrt{\frac{E_F}{E_0}} \right], \\
\sigma^z_{3D} &=& \frac{1}{\pi^2}\frac{e^2\tau}{\hbar^4} \frac{m_{xy}^*}{m_z^*} \sqrt{m^*_{xy} m_z^*} \alpha_R \left[ (E_F+E_0)\arcsin \left( \sqrt{\frac{E_0}{E_F+E_0}} \right)+\left( \frac{2}{3}E_F+E_0 \right)\sqrt{\frac{E_F}{E_0}} \right].
\end{eqnarray}
\end{widetext}
Like strong-exchange-coupling regime, only $E_{\parallel}$ contributes to the spin torque.
This is because the direction of spatial inversion symmetry breaking is $z$ axis, which implies the nature of 3D Rashba-type SOC.
In 2D Rashba models, the spin torque reads $\bm{T}_{2D}=-\frac{1}{2\pi}\frac{e\tau E_{\parallel}}{\hbar^4} m^*\alpha_R J_{sd}\cos(\varphi_M-\varphi_E) \bm{e}_z$ and the electric conductivity reads $\sigma^{\parallel}_{2D}=\frac{1}{\pi}\frac{e^2\tau}{\hbar^2} (E_F+E_0)$.

For low-density regime ($-E_0<E_F<0$), on the other hand, we obtain
\begin{eqnarray}
\bm{T}_{3D} &=& -\frac{1}{4\pi}\frac{e\tau}{\hbar^4}\sqrt{m^*_{xy} m_z^*} J_{sd}\left( E_F+E_0\right) \nonumber \\
 & & \times E_{\parallel} \cos(\varphi_M-\varphi_E) \bm{e}_z, \label{Ts2} \\
\sigma^{\parallel}_{3D} &=& \frac{1}{4\pi}\frac{e^2\tau}{\hbar^4} \sqrt{m^*_{xy} m_z^*} \alpha_R (E_F+E_0), \label{ss2} \\
\sigma^z_{3D} &=& \frac{2m_{xy}^*}{m_z^*}\sigma^{\parallel}_{3D}.
\end{eqnarray}
In 2D Rashba models, the spin torque is $\bm{T}_{2D}=-\frac{1}{\sqrt{2}\pi}\frac{e\tau}{\hbar^3} E_{\parallel}\sqrt{m^*} J_{sd}\sqrt{E_F+E_0} \cos(\varphi_M-\varphi_E) \bm{e}_z$ and the electric conductivity is $\sigma^{\parallel}_{2D}=\frac{1}{\sqrt{2}\pi}\frac{e^2\tau}{\hbar^3}\sqrt{m^*} \alpha_R \sqrt{E_F+E_0}$.

\subsection{Discussion}

\begin{figure}
 \begin{center} 
 \includegraphics[width=85mm]{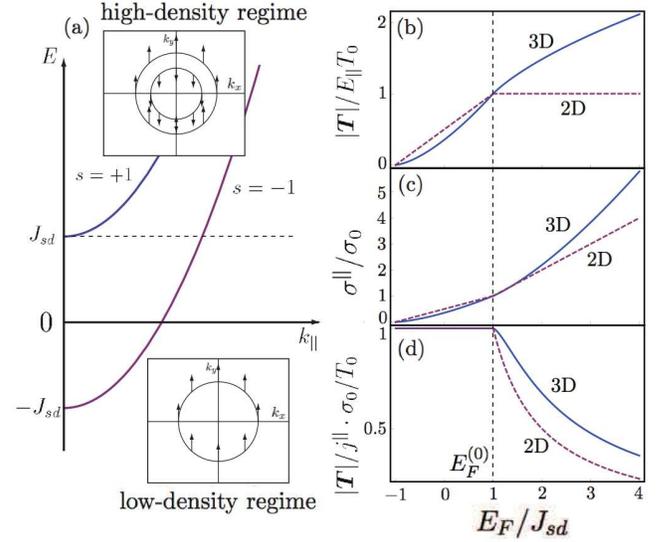} 
 \end{center} 
 \caption{(Color online) Strong exchange coupling regime of the 3D Rashba model. (a) Schematic of the band structure, (b) Fermi-energy dependence of the spin torque, (c) the electric conductivity and (d) the spin-torque efficiency. In each plot, the results for 2D Rashba models are shown for comparison. The dotted lines refer to a topological change of Fermi surfaces. The insets in (a) represent the Fermi surfaces and the spin density in the $k_x k_y$ plane ($k_z=0$) in low and high-density regimes. The arrows are parallel to the magnetization vector of the ferromagnet. Here the respective quantities are shown as ratios to these at the topological transition $E_F=J_{sd}$, i.e., $T_0^{3D}\equiv \frac{4}{3\pi^2}\frac{e\tau}{\hbar^5} m^*_{xy}\sqrt{m_z^*}\alpha_R J_{sd}^{3/2}$, $T_0^{2D}\equiv \frac{1}{\pi}\frac{e\tau}{\hbar^4} m^*\alpha_R J_{sd}$, $\sigma_{3D,0}^{\parallel}\equiv \frac{4}{3\pi^2} \frac{e^2\tau}{\hbar^3} \sqrt{m_z^*} J_{sd}^{3/2}$ and $\sigma_{2D,0}^{\parallel}\equiv \frac{1}{\pi}\frac{e^2\tau}{\hbar^2}J_{sd}$.} 
 \label{zee} 
 \end{figure}

\begin{figure}
 \begin{center} 
 \includegraphics[width=85mm]{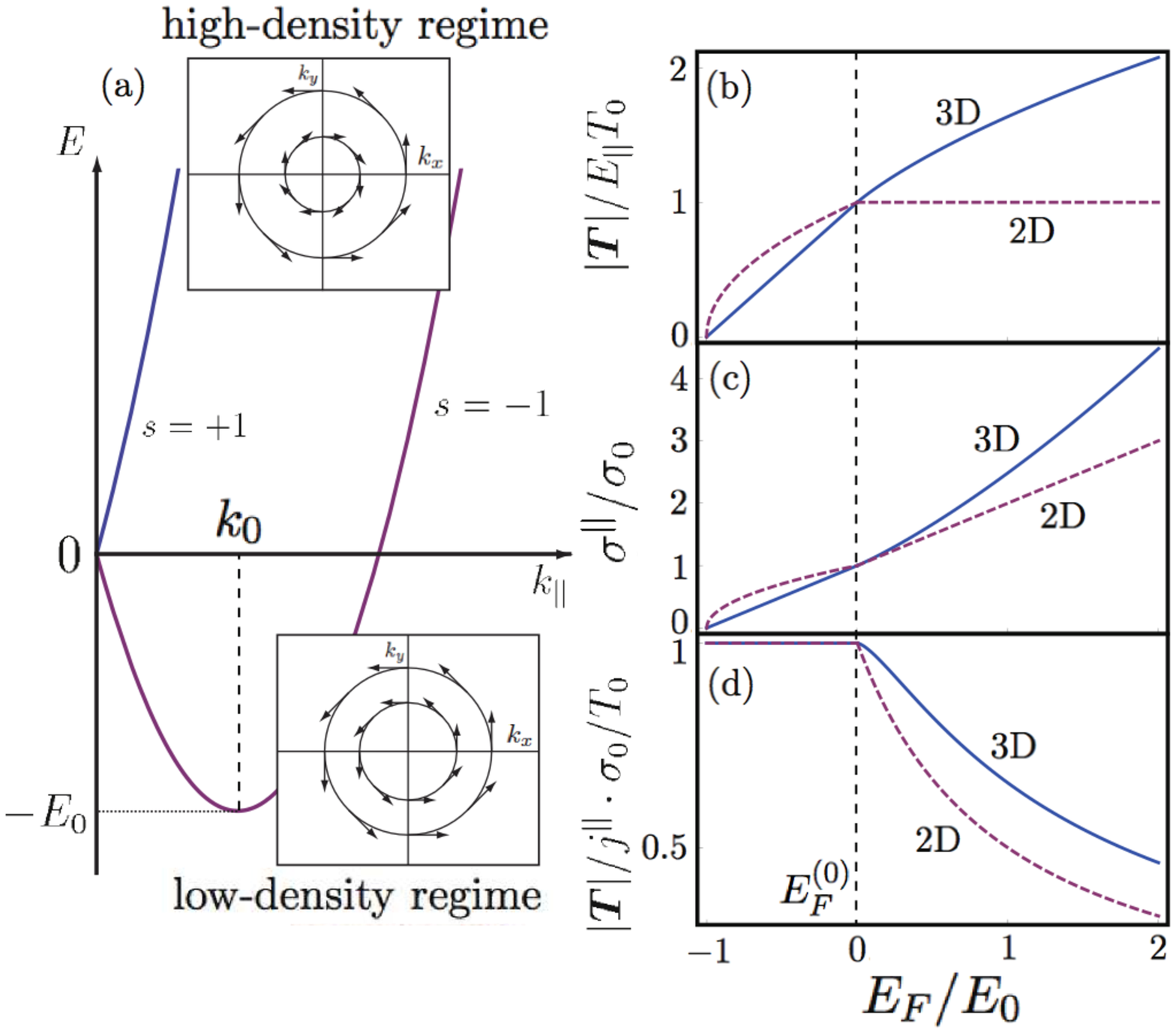} 
 \end{center} 
 \caption{(Color online) Weak exchange coupling regime of the 3D Rashba model. (a) Schematic of the band structure, (b) Fermi-energy dependence of the spin torque, (c) the electric conductivity and (d) the spin-torque efficiency. In each plot, the results for 2D Rashba models are shown for comparison. The dotted lines refer to a topological change of Fermi surfaces. The insets in (a) represent the Fermi surfaces and the spin density in the $k_x k_y$ plane ($k_z=0$) in low and high-density regimes. The arrows are along the Fermi surfaces. Here the respective quantities are shown as ratios to these at the topological transition $E_F=0$, i.e., $T_0^{3D}\equiv \frac{1}{4\pi}\frac{e\tau}{\hbar^4}\sqrt{m^*_{xy} m_z^*} J_{sd}E_0$, $T_0^{2D}\equiv \frac{1}{\sqrt{2}\pi}\frac{e\tau}{\hbar^3} \sqrt{m^*} J_{sd}\sqrt{E_0}$, $\sigma_{3D,0}^{\parallel}\equiv \frac{\sqrt{2}}{4\pi}\frac{e^2\tau}{\hbar^3} \sqrt{m_z^*} E_0^{3/2}$ and $\sigma_{2D,0}^{\parallel}\equiv \frac{1}{\pi}\frac{e^2\tau}{\hbar^2}E_0$.} 
 \label{rash} 
 \end{figure}

We show the Fermi-energy dependence of the spin torque, the electric conductivity and the spin-torque efficiency, defined as the ratio between the spin torque and the electric-current density, in (b), (c) and (d) of Figs. \ref{zee} and \ref{rash}.
Here we set $\varphi_M=\varphi_E$ for simplicity.
Let $E_F^{(0)}$ denote the value of the Fermi energy, where the topology of the Fermi surface changes.
This is the Fermi energy which differentiates between the low- and high-density regimes.
In these figures, we scale the respective quantities by their values at $E_F= E_F^{(0)}$.
As a result, each quantity is represented as a dimensionless value, which facilitates comparison between various cases.

As one can see for both limits, the slope of the spin torque shown in the low-density regime is suppressed in the high-density regime (Figs. \ref{zee} (b) and \ref{rash} (b)), while the slope of the electric conductivity in low-density regime is enhanced in the high-density regime (Figs. \ref{zee} (c) and \ref{rash} (c)).
The reason is because within our analysis, the spin torque is proportional to the spin-density component perpendicular to the magnetization and the spin distribution in the upper band is opposite to that in the lower band. 
Since the topology of the Fermi surface in the low-density regime differs from that in high-density regime, the behaviors are different between both regimes.
The spin torque as well as the electric conductivity is different in 3D and in 2D Rashba models due to the difference in dimensionality such as the density of states.
If we assume that $m^*_{xy}, \alpha_R, \tau$ in 3D are equal to $m^*, \alpha_R, \tau$ in 2D, the ratio between $|\bm{T}_{3D}|$ and $|\bm{T}_{2D}|$ in the low-density regime is proportional to the square root of $m^*_z$ and Fermi energy from the bottom of the conduction band.
This is asymptotically true in the limiting case of $E_F\gg J_{sd}, k_F\alpha_R$.

We also find that within the present approximation, the spin-torque efficiency $|\bm{T}|/j^{\parallel}$ is a constant in low-density regime (Figs. \ref{zee} (d) and \ref{rash} (d)).
In high-density regime, on the other hand, the spin-torque efficiency decreases monotonically both in 2D and in 3D.
For generic types of spin-split bands, we can draw analogy from the present simple models.
The spin-torque efficiency is expected to be larger, when only one of the spin-split bands lies at the Fermi energy.
When the Fermi energy becomes larger and crosses both of the spin-split bands, their contributions are expected to cancel partially.

Now we have assumed the constant relaxation-time approximation for both limiting regimes.
Since the relaxation time generally depends on the Fermi energy, the relaxation times $\tau$ in the expressions of the spin torque and the electric conductivity are replaced by $\tau_s(E_F)$.
However, the spin-torque efficiency is independent of the Fermi energy in the low-density regime because the spin torque and the electric current are both proportional to the relaxation time $\tau_-(E_F)$.
In particular, the spin-torque efficiency in the low-density regime does not alter for the above replacement $\tau \to \tau_s(E_F)$.
According to the formalism by Schliemann and Loss \cite{Schliemann03}, on the other hand, the effects of the anisotropic scattering due to the SOC is expressed in the distribution function:
\begin{eqnarray}
f_{\bm{k},s} &=& f^0_{\bm{k},s}+\frac{e}{\hbar} \frac{\tau^{\parallel}_{\bm{k},s}}{1+(\tau^{\parallel}_{\bm{k},s}/\tau^{\perp}_{\bm{k},s})^2} \bm{E}\cdot \frac{\partial f^0_{\bm{k},s}}{\partial \bm{k}} \nonumber \\
 & & +\frac{e}{\hbar} \frac{\tau^{\perp}_{\bm{k},s}}{1+(\tau^{\perp}_{\bm{k},s}/\tau^{\parallel}_{\bm{k},s})^2} (\bm{e}_z\times \bm{E})\cdot \frac{\partial f^0_{\bm{k},s}}{\partial \bm{k}},
\end{eqnarray}
where $\tau^{\parallel}_{\bm{k},s}$ and $\tau^{\perp}_{\bm{k},s}$ denote the longitudinal and the transverse relaxation time respectively.
For the strong exchange-coupling regime, the distribution function has a simple form $f_{\bm{k},s}=f^0_{\bm{k},s}+\frac{e}{\hbar}\tau_s(E_F) \bm{E}\cdot \frac{\partial f^0_{\bm{k},s}}{\partial \bm{k}}$ owing to an isotropic scattering probability.
For the weak exchange-coupling regime, the scattering probability becomes anisotropic, but $\tau^{\perp}_{\bm{k},s}$ vanishes within the zeroth order in $J_{sd}/k_F \alpha_R$, and the distribution function has a similar form with the case of isotropic scattering \cite{Manchon08}.
Therefore, we can ignore anisotropic scattering effects due to the SOC for both limits.

\section{Enhancement of spin-torque efficiency}

\begin{figure}
 \begin{center} 
 \includegraphics[width=85mm]{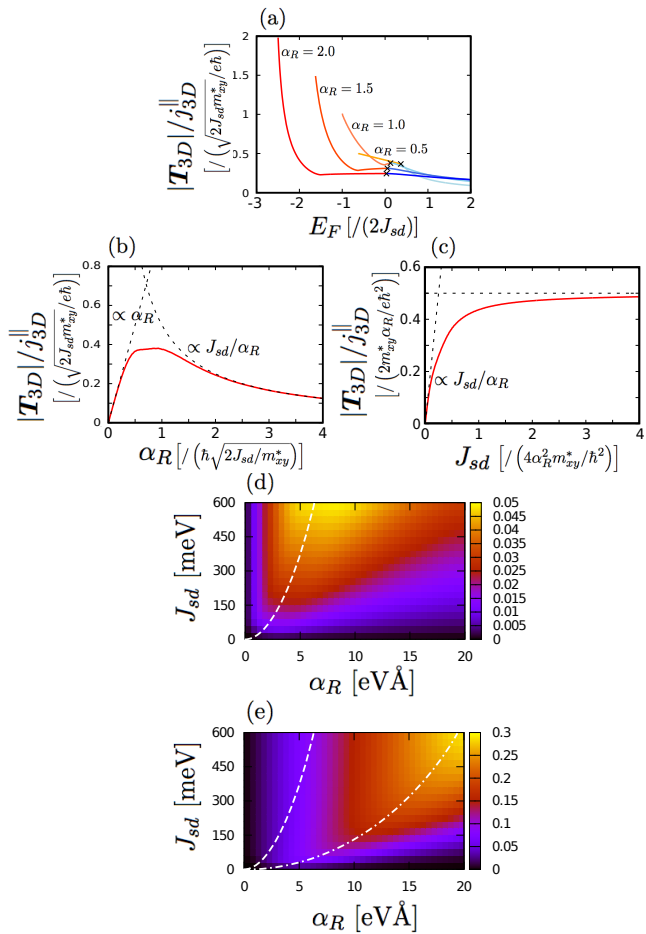} 
 \end{center} 
 \caption{(Color online) Numerical results for the spin-torque efficiency.
 (a) Spin-torque efficiency for various values of $\alpha_R$ as a function of $E_F$.
 The crosses denote topological changes of the Fermi surfaces.
 (b) Spin-torque efficiency at the topological transition as a function of $\alpha_R$, and (c) that as a function of $J_{sd}$.
 (d) Spin-torque efficiency as a function of $\alpha_R$ and $J_{sd}$ at the topological transition, and (e) that at the band bottom.
 In (d) and (e), we adopted $m^*_{xy}/\hbar^2=0.014 \ \mathrm{/eV\AA^2}$, $m^*_z/m^*_{xy}=0.5$ and $\varphi_E=\varphi_M$.
 The spin-torque efficiencies in (d)-(e) are plotted as a unit of $e$.
 The broken curves in (b) (c) represent the asymptotic forms of the spin-torque efficiency for $J_{sd}\gg k_F\alpha_R$ and $k_F\alpha_R\gg J_{sd}$.
 The white broken curves in (d) (e) represent the optimal condition $2E_0\sim J_{sd}$ and the dot-dashed curve in (e) denotes the modified optimal condition $2E_0\sim 9J_{sd}$.} 
 \label{tre} 
 \end{figure}

To realize magnetization-switching devices, it is indispensable to enhance the spin-torque efficiency in order to minimize the threshold electric current for the magnetization reversal.
So far we have found that the spin-torque efficiency in the low-density regime is a constant, both for weak and strong exchange coupling regime.
On the other hand, by numerical analysis in the intermediate exchange coupling regime, the spin-torque efficiency as a function of $E_F$ is not a constant in the low-density regime.
As shown in Fig. \ref{tre} (a), when $E_F$ goes to the band bottom, the spin-torque efficiency is largely enhanced in a non-monotonic fashion.
The spin-torque efficiencies in the low-density regime also depend on $m^*_{xy}$, $\alpha_R$ and $J_{sd}$.
Let us consider how to optimize the spin-torque efficiency by varying $\alpha_R$.
From Eqs. (\ref{Ts1}) and (\ref{ss1}) for $J_{sd}\gg k_F\alpha_R$, and Eqs. (\ref{Ts2}) and (\ref{ss2}) for $k_F\alpha_R\gg J_{sd}$, the spin-torque efficiencies read for the low-density regime
\begin{eqnarray}
|\bm{T}_{3D}|/j^{\parallel}_{3D} &=& \frac{m_{xy}^*}{e\hbar^2}\alpha_R \ :J_{sd}\gg k_F\alpha_R, \label{Tj1} \\
|\bm{T}_{3D}|/j^{\parallel}_{3D} &=& \frac{1}{e} \frac{J_{sd}}{\alpha_R} \ :k_F\alpha_R\gg J_{sd}. \label{Tj2}
\end{eqnarray}
When $J_{sd}$ is kept constant and $\alpha_R$ is varied, the spin-torque efficiency increases linearly in $\alpha_R$ for $J_{sd}\gg k_F\alpha_R$ and decreases inversely linear in $\alpha_R$ for $k_F\alpha_R\gg J_{sd}$.
Thus, it implies that the spin-torque efficiency becomes maximum at an intermediate value of $\alpha_R$. 
From these asymptotics (\ref{Tj1}) and (\ref{Tj2}), we find that the optimal condition is expected to be $\alpha_R\sim \hbar \sqrt{J_{sd}/m^*_{xy}}$, i.e., $2E_0\sim J_{sd}$.
Namely, the spin-splitting energy $E_0$ by the SOC is comparable to the exchange energy $J_{sd}$ with the localized spins.
The maximum spin-torque efficiency is $k_0/e$, proportional to the Rashba momentum.
To confirm our expectations, we show the $\alpha_R$ dependence of the spin-torque efficiency at the topological transition $E_F=E_F^{(0)}$ in Fig. \ref{tre} (b).
The spin-torque efficiency becomes maximum near $\alpha_R\sim \hbar \sqrt{J_{sd}/m^*_{xy}}$, which confirms our expectation.
When $\alpha_R$ is kept constant and $J_{sd}$ is varied, on the other hand, the spin-torque efficiency increases linearly in $J_{sd}$ for $k_F\alpha_R\gg J_{sd}$ and goes to a constant $k_0/e$ for $J_{sd}\gg k_F\alpha_R$ as shown in Fig. \ref{tre} (c).
Using realistic parameters, we numerically show the spin-torque efficiency as a function of $\alpha_R$ and $J_{sd}$ at the topological transition and at the band bottom in Fig. \ref{tre} (d) and (e) respectively.
At the band bottom, the broken curve in Fig. \ref{tre} (e) shows that the optimal condition is largely shifted to the larger value of $\alpha_R$, roughly given by $2E_0\sim 9J_{sd}$.
This shift comes from the large enhancement of spin-torque efficiency at the band bottom for larger $\alpha_R$, shown in Fig. \ref{tre} (a).

\begin{figure}[H]
 \begin{center} 
 \includegraphics[width=60mm]{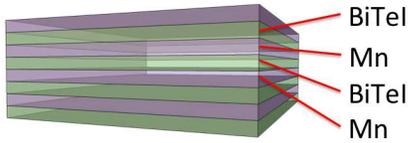} 
 \end{center} 
 \caption{(Color online) Candidate geometry for the magnetization reversal by spin torque from 3D Rashba system BiTeI.
 Bulk BiTeI is sandwiched between ferromagnets Mn, similar to a Rashba semiconductor doped with a ferromagnet.
 } 
 \label{mnbite} 
 \end{figure}

Finally we suggest methods to realize the magnetization reversal in 3D Rashba systems experimentally.
One way is to synthesize ferromagnetic bulk Rashba semiconductors by doping magnetic impurities into bulk materials with a strong SOC, e.g., BiTeI.
Nevertheless, synthesis of a new ferromagnetic semiconductor is quite difficult, which would be a challenging and promising issue for materials science.
On the other hand, it is known that wide-gap semiconductors such as Mn-doped GaN, Co-doped ZnO and V-doped ZnO present ferromagnetism even over the room temperature \cite{Dietl00, Saeki01, Sonoda02}.
GaN and ZnO have wurtzite-type crystalline structure and are conventional Rashba semiconductors.
Since these semiconductors present much smaller Rashba spin splitting (about $\alpha_R=1.1$ meV$\AA$ in ZnO and $\alpha_R=9$ meV$\AA$ in GaN) \cite{Voon96, Majewski05}, BiTeI is suitable for demonstrating the magnetization reversal to fabricate ferromagnetic Rashba semiconductors.
Another way is to fabricate layered materials with a Rashba semiconductor film and a ferromagnetic metal as shown in Fig. \ref{mnbite}.
These materials are considered to be similar to a magnetically doped Rashba semiconductor macroscopically.
We also remark that in conventional ferromagnetic 2DEGs, only the magnetization near the interface is switched, while in ferromagnetic 3D Rashba semiconductors, the magnetization of the whole crystal is switched.

Let us evaluate the physical quantities using realistic parameters.
We take BiTeI doped with magnetic element as an example, and use the values $\alpha_R=3.85 \ \mathrm{eV \AA}$ and $m^*_{xy}/\hbar^2=0.014 \ \mathrm{/eV \AA^2}$ for BiTeI \cite{Ishizaka11}.
For numerical estimate, we assume Mn as dopant, and take the saturation magnetization $M_s=10^4 \ \mathrm{J/T\cdot m^3}$ and the anisotropy field $H_K=200 \ \mathrm{Oe}$ adopted from Mn-doped semiconductors \cite{Manchon08}.
The estimated maximum spin-torque efficiency is $k_0/e\sim 3\times 10^{27}\ \mathrm{/C\cdot m}$ for the low-density regime, which is realized for the optimal condition $J_{sd}\geq 20\ \mathrm{meV}$ if the Fermi energy lies in the vicinity of the band bottom.
Therefore, the critical electric current density is evaluated as $j_c\sim 6\times 10^4\ \mathrm{A/cm^2}$, which is much lower than that observed in the 2D Rashba system such as Pt/Co/AlO$_x$ junction \cite{Miron10}.

\section{Summary}

We have investigated a spin torque and its efficiency induced by the Rashba spin-orbit coupling in three dimensions using the Boltzmann transport theory.
It was shown that in the high-density regime, the increase of the spin torque as a function of the Fermi energy is slower than that in the low-density regime.
It is because in the high-density regime, there occurs cancellation between the two spin-split bands.
We also found that high spin-torque efficiency is achieved when the Fermi energy lies on only the lower band and there exists an optimal value for the Rashba parameter.
The spin-torque efficiency becomes maximum when the Rashba spin-splitting energy is comparable to the exchange energy with the localized spins, and then its maximum values are determined by the Rashba momentum.
Such optimization might be useful for the magnetization reversal.

\begin{acknowledgments}
This work was supported by Grant-in-Aids from MEXT, Japan (No. 21000004 and No. 22540327) and the Kurata Grant from The Kurata Memorial Hitachi Science and Technology Foundation.
\end{acknowledgments}


\begin{thebibliography}{99}
 \bibitem{Slonczewski96}J. C. Slonczewski, J. Magn. Magn. Mater. \textbf{159}, L1 (1996).
 \bibitem{Berger96}L. Berger, Phys. Rev. B \textbf{54}, 9353 (1996).
 \bibitem{Tan07}S. G. Tan, M. B. A. Jalil, X. J. Liu, arXiv:0705.3502.
 \bibitem{Obata08}K. Obata and G. Tatara, Phys. Rev. B \textbf{77}, 214429 (2008).
 \bibitem{Manchon08}A. Manchon and S. Zhang, Phys. Rev. B \textbf{78}, 212405 (2008).
 \bibitem{Manchon09}A. Manchon and S. Zhang, Phys. Rev. B \textbf{79}, 094422 (2009).
 \bibitem{Matos-Abiague09}A. Matos-Abiague and R. L. Rodruez-Suez, Phys. Rev. B \textbf{80}, 094424 (2009).
 \bibitem{Tan11}S. G. Tan, M. B. A. Jalil, T. Fujita, X. J. Liu, Ann. Phys. \textbf{326}, 207 (2011).
 \bibitem{Chernyshov09}A. Chernyshov, M. Overby, X. Liu, J. K. Furdyna, Y. Lyanda-Geller, and L. P. Rokhinson, Nature Phys. \textbf{5}, 656 (2009).
 \bibitem{Miron10}I. M. Miron, G. Gaudin, S. Auffret, B. Rodmacq, A. Schuhl, S. Pizzini, J. Vogel, and P. Gambardella, Nature Mater. \textbf{9}, 230 (2010).
 \bibitem{Rashba60}E. I. Rashba, Sov. Phys. Solid State \textbf{2}, 1109 (1960); Yu. A. Bychkov and E. I. Rashba, J. Phys. C \textbf{17}, 6039 (1984).
 \bibitem{Ishizaka11}K. Ishizaka, M. S. Bahramy, H. Murakawa, M. Sakano, T. Shimojima, T. Sonobe, K. Koizumi, S. Shin, H. Miyahara, A. Kimura, K. Miyamoto, T. Okuda, H. Namatame, M. Taniguchi, R. Arita, N. Nagaosa, K. Kobayashi, Y. Murakami, R. Kumai, Y. Kaneko, Y. Onose, and Y. Tokura, Nature Mater. \textbf{10}, 521 (2011).
 \bibitem{Bahramy11}M. S. Bahramy, R. Arita, and N. Nagaosa, Phys. Rev. B \textbf{84}, 041202(R) (2011).
 \bibitem{Nitta97}J. Nitta, T. Akazaki, H. Takayanagi, and T. Enoki, Phys. Rev. Lett. \textbf{78}, 1335 (1997).
 \bibitem{LaShell96}S. LaShell, B. A. McDougall, and E. Jensen, Phys. Rev. Lett. \textbf{77}, 3419 (1996).
 \bibitem{Ast07}C. R. Ast, J. Henk, A. Ernst, L. Moreschini, M. C. Falub, D. Pacile, P. Bruno, K. Kern, and M. Grioni, Phys. Rev. Lett. \textbf{98}, 186807 (2007).
 \bibitem{Lifshitz60}I. M. Lifshitz, Sov. Phys. -JETP \textbf{11}, 1130 (1960).
 \bibitem{Schliemann03}J. Schliemann and D. Loss, Phys. Rev. B \textbf{68}, 165311 (2003).
 \bibitem{Dietl00}T. Dietl, H. Ohno, F. Matsukura, J. Cibert, and D. Ferrand, Science \textbf{287}, 1019 (2000).
 \bibitem{Saeki01}H. Saeki, H. Tabata, and T. Kawai, Solid State Commun. \textbf{120}, 439 (2001).
 \bibitem{Sonoda02}S. Sonoda, S. Shimizu, T. Sasaki, Y. Yamamoto, and H. Hori, J. Cryst. Growth. \textbf{237}, 1358 (2002).
 \bibitem{Voon96}L. C. Lew Yan Voon, M. Willatzen, M. Cardona, and N. E. Christensen, Phys. Rev. B \textbf{53}, 10703 (1996).
 \bibitem{Majewski05}J. A. Majewski and P. Vogl, Physics of Semiconductors: 27th International Conference on the Physics of Semiconductors, edited by J. Menendez and C. G. Van de Walle (American Institute of Physics, 2005), p.1403.
\end{thebibliography}
\end{document}